\documentclass[]{article}
\usepackage{amsmath,graphicx,full page,threeparttable}
\usepackage[titletoc]{appendix}
\newcommand{\tbl}[1] {\caption{#1}  }
\newcommand{\colrule}[0]{\hline}
\newcommand{\botrule}[0]{ \bottomrule }
\newcommand{\tabnote}[1] {\item #1}

\usepackage{multirow,booktabs,url}
\usepackage[stable]{footmisc}

\begin{document}

% TITLE PAGE
%\setcounter{page}{0}
%\begin{Large}
%\begin{center}
%ARTICLE
%\vskip 5cm
%{\bf A modified ziggurat algorithm for generating exponentially- and normally-distributed pseudorandom numbers} \\
%\end{center}
%\vskip 2cm
%Christopher D McFarland$^1$	\\

%\vskip 1cm
%\noindent
%$^1$ Graduate Program in Biophysics, Harvard University, Boston, MA 02115 \\
%\vskip 2cm
%\noindent
%Corresponding author: Christopher D McFarland mcfarlan@fas.harvard.edu, E25-524, MIT, 77 Massachusetts Ave, Cambridge, MA 02139, (617) 452-4075
%\end{Large}
%\clearpage

% JOURNAL SPECIFIC HEADER
%\jvol{00} \jnum{00} \jyear{2014} \jmonth{January}

%\articletype{ARTICLE}

% TRUE BEGINNING

\title{A modified ziggurat algorithm for generating exponentially- and normally-distributed pseudorandom numbers}

\author{Christopher D McFarland$^{\rm a \ast}$
\thanks{$^\ast$
Corresponding author. Email: mcfarlan@fas.harvard.edu}
\\\vspace{6pt}  $^{a}${\em{Program in Biophysics, Harvard University, Cambridge, Massachusetts, USA}}}

\maketitle

\begin{abstract}
The Ziggurat Algorithm is a very fast rejection sampling method for generating PseudoRandom Numbers (PRNs) from common statistical distributions. The algorithm divides a distribution into rectangular layers that stack on top of each other (resembling a Ziggurat), subsuming the desired distribution. Random values within these rectangular layers are then sampled by rejection. This implementation splits layers into two types: those constituting the majority that fall completely under the distribution and can be sampled extremely fast without a rejection test, and a few additional layers that encapsulate the fringe of the distribution and require a rejection test. This method offers speedups of 65\% for exponentially- and 82\% for normally-distributed PRNs when compared to the best available C implementations of these generators. Even greater speedups are obtained when the algorithm is extended to the Python and MATLAB/OCTAVE programming  environments.
%\begin{keywords}
%Ziggurat Method, Pseudorandom numbers, C, Python, Matlab, exponential distribution, normal distribution
%\end{keywords}
\end{abstract}

\section{Introduction}
Random numbers are essential for a variety of applications: the modeling of natural systems, optimization, and cryptography, to name a few. However, computers are designed to behave deterministically, thus making truly random number generation from a computer often difficult, and sometimes impossible. PseudoRandom Numbers (PRNs), or deterministic random numbers, are generally used as a reasonable substitute for truly random numbers. Because of their wide-range of applications, PRNs have a long history of study. PRN Generators (PRNGs) most often work by transforming an initial single random number, or \lq{}seed\rq{}, into a new PRN, and then using the new PRN to seed a transformation into the next PRN. While the transformation algorithm in PRNGs is deterministic, it nevertheless satisfies important properties of truly random numbers, such as large periodicity, equidistribution and discontinuity \cite{Lecuyerl1992}. Most current PRNGs output uniformly-distributed values. These uniformly-distributed PRNs are then transformed into other sampling distributions by downstream algorithms. Often, this transformation takes significantly greater time than the initial uniform PRNG, thus constituting the primary bottleneck of some stochastic algorithms. 

The Ziggurat Algorithm is the most commonly used method to obtain non-uniformly-distributed PRNs. It was first proposed in the early 60's \cite{Marsaglia1984} and has since been modified many times \cite{Marsaglia2000, Doornik2005, Zhang2005}, currently being among the fastest methods available on modern CPUs \cite{Marsaglia2000}, although other fast methods exist \cite{Rubin2006}. The algorithm works via rejection sampling, a three-step process for generating random numbers. (1) The desired probability distribution $P(x)$ is subsumed by a set of boxes, resembling a ziggurat. The design of these boxes is described below. (2) Two uniform PRNs are used to define a point $(x, y)$ within a randomly chosen box. (3) If this point lies beneath the desired probability distribution, i.e. if $y < P(x)$, then the $x$ coordinate is returned; otherwise the point is `rejected' and a new point $(x, y)$ is selected and tested. 

Here, we present a modified Ziggurat Algorithm that creates rectangular layers that lie completely beneath $P(x)$, rather than completely containing $P(x)$. This eliminates the need to sample these layers by rejection, but also leaves short gaps of probability mass that must be sampled in a small minority of iterations. By eliminating the need to rejection sample most PRNs and by sampling these small gaps of probability mass efficiently, exponentially- and normally-distributed PRN generation is greatly accelerated. In the next section, the modified algorithm is described in detail alongside the traditional ziggurat method. I then discuss timings of the algorithm in comparison to the best alternative algorithms and demonstrate a considerable speedup. In the appendix, I present the code, affirm the random properties of the generated distributions, and discuss additional minor optimizations that further improved performance.

\section{Description of the algorithm}\label{description}
As detailed above, a uniform PRNG is utilized as an input source of randomness for the Ziggurat method. Here, a popular Mersenne Twister algorithm \cite{Saito2008} is used to generate uniform PRNs. This Mersenne Twister runs very fast and exhibits excellent randomness, making it ideal for use in most applications excluding cryptography. Nevertheless, the generator can be seamlessly substituted. 

In a ziggurat algorithm, the desired probability distribution $P(x)$ lies beneath a stack of rectangular layers. Layers are designed such that they all contain the exact same area, so that each box can be randomly chosen with equal probability using a uniform random integer $i$ to ensure uniform coverage of $P(x)$. The height $f_i$ and length $X_i$ of each ziggurat layer are pre-calculated in lookup tables. Because of these  lookup tables, ziggurat algorithms are most efficiently implemented on systems with large caches (e.g. modern CPUs, but not current GPUs) \cite{Thomas2007}. 

Ziggurat algorithms accelerate computation because the vast majority of points within the ziggurat layers reside in regions that are a priori guaranteed to lie beneath $P(x)$ \cite{Marsaglia2000, Zhang2005} (Figure 1). By avoiding sampling by rejection the algorithm is greatly accelerated since most probability distributions are transcendental and, thus, require significant time to calculate $P(x)$. In the traditional exponentially-distributed ziggurat algorithm, greater than 3\% of the distribution will be rejection tested when $i_{\rm{max}} = 256$ ziggurat layers are used \cite{Marsaglia2000}. 

$P(x)$ often contains a tail that resides outside of the ziggurat layers. Sampling from this tail can always be achieved via Inverse Transform Sampling \cite{DeSchryver2010}. However, for certain probability distributions faster approaches are possible. In general, the ziggurat algorithm is ideal for distributions where sampling from the tail is rare. 

The modified ziggurat algorithm presented here differs from the traditional algorithm in one key manner: layers lie completely beneath $P(x)$, whereas layers completely subsume $P(x)$ in the traditional algorithm (Figure 1). This modification eliminates the need for a rejection test within the ziggurat layers, however it leaves small gaps of probability mass to the right of each layer. These overhanging gaps are then sampled in a small minority of cases via an efficient algorithm that I will describe below. 

\begin{figure}
\begin{center}
\includegraphics[width=\textwidth]{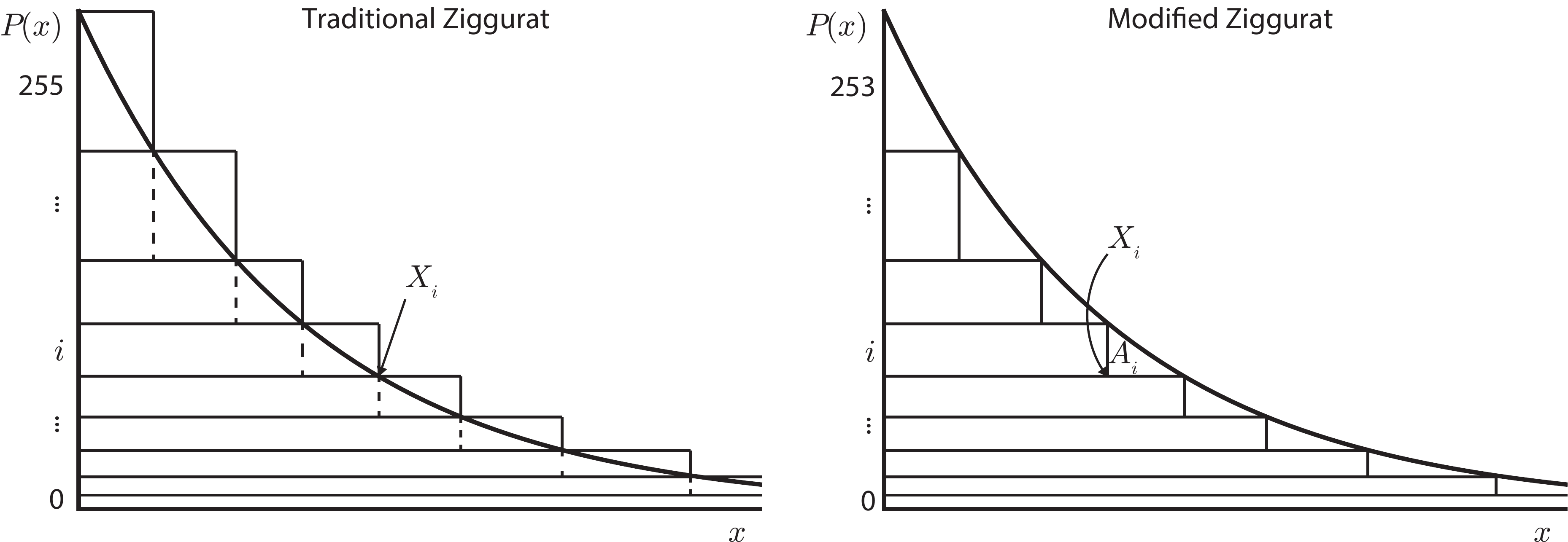}
\caption{\label{Fig1} \textbf{Ziggurat layers in the modified algorithm lie completely beneath $P(x)$.} 
In the Traditional Ziggurat, layers completely contain the desired distribution (excluding the tail). Because these layers are rectangular, they must extend beyond $P(x)$, thus requiring a rejection sampling test in a subset of cases. In the proposed algorithm, layers reside completely beneath $P(x)$. This eliminates the need for a rejection test when sampling from the ziggurat layers. However, small gaps of probability mass, with area $A_i$, overhang to the right of each layer. These gaps must be sampled in $<2\%$ of cases, using a rejection test described later.}
\label{Figure1}
\end{center}
\end{figure}

To lie completely beneath the desired distribution, ziggurat layers must extend until their \emph{upper-right} corner coincides with $P(x)$ (traditionally, their \emph{lower-right} corner coincides with $P(x)$). The position of this corner is then $\left( X_i,\  f_i = P(X_i) \right)$, where $X_i$ is the length of each layer. Like the traditional ziggurat algorithm, the lower-left corner $\left(0, f_{i-1} = P(X_{i-1}) \right)$ begins at $x = 0$ and lies immediately above the previous layer.  Also like the traditional algorithm, layers are equal in area. 
However, the area of each layer is now slightly smaller. In the new algorithm, a PRN is always returned from each layer when it is selected. Thus, its area \emph{must} be exactly $1/i_{\rm{max}}$. In the traditional algorithm, PRNs are sometimes rejected, so their areas are slightly larger). 

With this constraint on each layer\rq{}s volume, we can solve for $X_i$:
\begin{equation*}
1/i_{\rm{max}} = X_i \left( P(X_i) - P(X_{i-1}) \right)
\end{equation*}
This iterative equation is solvable numerically using the Bisection Method. The first layer begins with its lower-left corner at the origin $(0, 0)$, and subsequent layers are continually solved until no more layers can be created. Small un-sampled overhangs of probability mass of area $A_i = \int_{X_i}^{X_{i-1}} P(x) - P(X_{i-1}) dx$ remain to the right of each layer (Figure 1). 

Less than $i_{\rm{max}}$ rectangular layers will fit beneath $P(x)$, as each layer is exactly $1/i_{\rm{max}}$ in area, yet additional overhangs remain. Indeed, the total number of layers in the modified algorihtm $L_{\rm{max}}$ cannot be determined until the last layer is calculated, which for an exponential distribution $L_{\rm{max}} = 252$ when $i_{\rm{max}} = 256$ (in the \emph{Appedix}, I show that 256 is optimal among the values of $i_{\rm{max}}$ tested). Thus, the probability mass overhangs in the exponential case consume $4/256 = 1.6\%$ of the total volume.

Like the traditional ziggurat algorithm, the modified algorithm relies on 3 pre-calculated tables. In the modified algorithm, these tables are the lengths of each ziggurat layer $X_i$, the height of each layer $f_i = P(X_i)$, and the area of each gap to the right of each layer $A_i$. Both algorithms also rely upon uniform floating-point PRNs $U_1, U_2 \in [0, 1)$ and a uniform integer PRN $i \in [0, i_{\rm{max}})$. For the modified ziggurat algorithm, sampling from the overhangs requires an additional PRN integer $j \in [0, L_{\rm{max}} )$, which is sampled from a non-uniform discrete distribution defined by the probability mass vector $A$. This sampling is accomplished in $\mathcal{O}(1)$ operations using a previously-described algorithm \cite{Smith2002}.

Table 1 describes in pseudocode the modified ziggurat algorithm alongside the traditional algorithm. In the modified algorithm, if the rectangle chosen is less than $L_{\rm{max}}$, then $x$ is immediately drawn and returned---eliminating several operations. For this reason, and because the exceptional case (i.e. progression to the end of the algorithm) is less common in the modified algorithm, it is faster. 

\begin{table}
\centering
\tbl{Comparison of modified and traditional ziggurat algorithms}
{\begin{tabular}[l]{@{}p{6cm}p{6cm}}
\toprule
\multicolumn{1}{c}{\textbf{Modified algorithm}} & \multicolumn{1}{c}{\textbf{Traditional algorithm}}		\\
\colrule
1. Generate $U_1$, $i$						& 1. Generate $U_1$, $i$						\\
2. If $i < L_{\rm{max}}$, return $U_1 X_i$			& 2. $x \leftarrow U_1 X_i$		\\
3. Generate $j$	from $A$						& 3. If $U_1 < k_i$, return $x$				\\
4. If $j = 0$, return a value from the tail		& 4. If $i = 0$, return a value from the tail		\\
5. Generate $U_2$							& 5. Generate $U_2$							\\
6. $x \leftarrow X_j + U_1 (X_{j-1} - X_j)$		& 6. If $(f_{i-1} - f_i ) U_2 < P(x)$, return $x$	\\
7. If $(f_{i-1} - f_i ) U_2 < P(x)$, return $x$		& 7. Go to 1.								\\
8. Go to 4.								& \\
\hline
\multicolumn{2}{c}{\textbf{Operations executed in the common case}} \\
\hline
\multicolumn{1}{c}{\textbf{Modified algorithm}}	& \multicolumn{1}{c}{\textbf{Traditional algorithm}}	\\
\multicolumn{1}{c}{\emph{98.4\% probability of exit at step 2.}}	& \multicolumn{1}{c}{\emph{97.8\% probability of exit at step 3.}} \\ 
1. Generate $U_1$					& 1. Generate $U_1$		\\
2. Generate $i$ 					& 2. Generate $i$ 		\\
3. Compare $i < L_{\rm{max}}$ 			& 3. Lookup $X_i$		\\ 
4. Lookup $X_i$					& 4. Multiply $U_1 X_i$	\\
5. Multiply $U_1 X_i$				& 5. Assign $x$			\\
								& 6. Lookup $k_i$		\\
								& 7. Compare $U_1 < k_i$	\\	
\botrule
\end{tabular}}
\label{Table1}
\end{table}

Additional modifications, exploiting the mathematical properties of normal and exponential distributions, can be made to accelerate sampling in the exceptional case (steps 3-8 in Table 1) where points are rejection sampled or sampled from the tail. Sampling from the tail can be accelerated by noting that the exponential distribution is memoryless, i.e. the tail of an exponential distribution is, itself, an exponential distribution \cite{Rubin2006}. Hence, values from the tail can be drawn using the ziggurat algorithm recursively. For the normal distribution, a previously described algorithm that transforms exponentially-distributed PRNs accelerates sampling from the tail \cite{Marsaglia2000}. 

Lastly, rejection sampling can be avoided in most cases even when sampling from the overhanging boxes in an exponential distribution. These boxes can be split into three subspaces: (i) a triangular area exclusively above $P(x)$---note that the exponential distribution has negative curvature everywhere, so any line segment between two points on $P(x)=e^{-x}$ lies completely above $P(x)$; (ii) a triangular area exclusively below $P(x)$, and (iii) a narrow band of area, proximal to the $P(x)$ curve that must still be sampled by rejection (Figure 2). The upper bound for this narrow band is simply the line segment connecting the points ($X_i$, $f_i = P(X_i)$) and ($X_{i+1}$, $f_{i+1}$), which is $y = f_i + (x - X_i)(f_{i-1} -f_i)$. The lower bound is defined by considering the maximum deviation $\epsilon$ of $P(x)$ from this upper bound: 
\vspace{-8pt}
\begin{flalign*}
& \epsilon = \mbox{max}_x \left[ f_i + (x - X_i)(f_{i-1} - f_i) - P(x) \right] 			\\
& \epsilon = f_{i-1} + \left( \mbox{Log}(f_i - f{i-1}) + X_i \right) \left( f_i - f_{i-1} \right)
\end{flalign*}
Because an exponential distribution is nearly linear over short distances, this deviation is quite small. When $i_{\rm{max}} = 1/256$, the widest narrow band is still only 9\% of the ziggurat box height. Hence, partitioning the overhang boxes into 3 regions eliminates 91\% of all rejection tests, further accelerating the algorithm. A few additional incremental speedups are described in the \emph{Appendix}. 

\begin{figure}
\begin{center}
\includegraphics[width=0.8\textwidth]{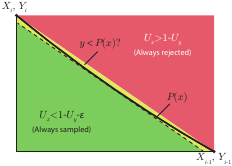}
\caption{\label{Fig2} \textbf{Rejection sampling of ziggurat overhangs in a exponential distribution can be further accelerated.} 
Consider the overhanging probability masses from Figure 1. Random points ($U_x$, $U_y$) within these smaller overhang boxes are sampled by rejection: points below $P(x)$ are returned, while points above $P(x)$ are rejected. Most rejection tests in these overhangs are avoided, further accelerating computation, by partition the overhang boxes into 3 sections: an area where sampling never succeeds ($U_x > 1 - U_y$), an area where sampling always succeeds ($U_x > 1 - U_y - \epsilon$), and a small narrow band proximal to $P(x)$ ($U_y - U_x < \epsilon$), where rejection tests are still necessary.}
\label{Figure2}
\end{center}
\end{figure}

\section{Implementation}
The algorithm was originally implemented in C and then embedded in Python and MATLAB/Octave using wrapper functions that mimic behavior of native functions (see \emph{Appendix} for source code). Lookup tables were calculated in a separate script and then inserted directly into the source code of the C implementation. The uniform PRNG described previously \cite{Saito2008} generates an array of uniform PRNs to capitalize on SIMD instructions and maximize speed. I made slight modifications to this code that minimized index checking, minimized function calls, deprecated support for old architectures not supported by this algorithm, and automatically seeds the PRNG using the system time, process ID, and parent process ID. Source code was designed such that this uniform PRNG can be easily substituted.

\section{Timings}
\begin{figure}
\begin{center}
\includegraphics[width=0.5\textwidth]{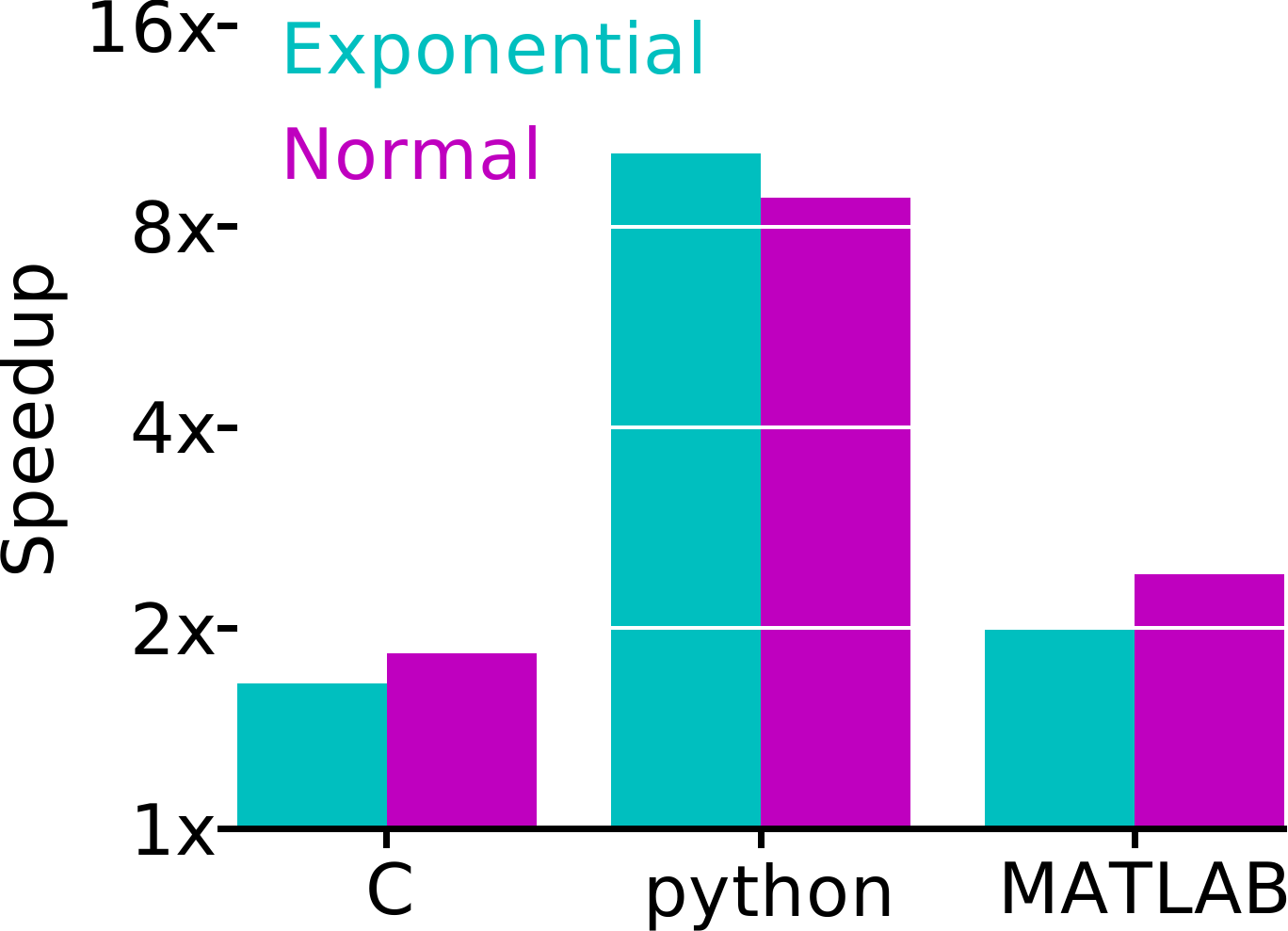}
\caption{\label{Fig3} \textbf{The modified ziggurat algorithm outperforms all other algorithms.} 
Speedup ranged from 65\% to $>1,000$\%, although the most impressive gains occur in programming environments where performance is prioritized less. Functions that mimic native PRNGs in python and MATLAB/Octave are provided, allowing seamless installation and integration (see \emph{Appendix}).}
\label{Figure3}
\end{center}
\end{figure}

The modified ziggurat outperforms all other exponentially- and normally-distributed PRNGs. The speedup, or timing of the fastest alternative algorithm divided by this algorithm's speed, was 65\% or greater for the various programming languages tested (Figure \ref{Figure3}). In the comparison, the median runtime of three trials of generating and aggregating $10^9$ PRNs on two different architectures (circa 2012) are presented (Table 2). In short, the fastest C implementation of this algorithm generates uniform PRNs, transforms these values into exponentially-distributed PRNs, and adds these values to an aggregate sum in $<10$ CPU cycles per iteration. 

\begin{table}
\centering
\tbl{Performance of the modified ziggurat algorithm across architectures.}
{\begin{tabular}[l]{@{}lccc@{}}
\toprule
Algorithm	& Architecture 1$^{\rm a, b}$ (s)	& Architecture 2$^{\rm c}$ (s) & Average Speedup	\\
\hline
exponential.h			&	2.79	&	3.37 	&	\multirow{2}{*}{1.65}	\\
Marsaglia \& Tsang \cite{Marsaglia2000} 	&	4.63	&	5.56	& 	\vspace{5pt} 		\\
normal.h				&	3.33	&	4.03	&	\multirow{2}{*}{1.83}	\\
Doornik \cite{Doornik2005}			& 	6.19	&	7.24	&	\vspace{5pt}		\\	
fast\_prns:exponential	&	4.15	&	5.05	&	\multirow{2}{*}{10.3\: \: }	\\
numpy:exponential$^{\rm d}$ 		&	42.8	\: \: &	52.1	\: \: &	\vspace{5pt}	\\	
fas\_prns:normal		&	4.42	&	5.05	&	\multirow{2}{*}{8.85}		\\
numpy:normal			&	37.8 \: \:  &	46.1	\: \: &	\vspace{5pt}	\\	
cdm\_exprnd				&	7.40	&	8.73	&	\multirow{2}{*}{1.99}	\\
Matlab R2013a exprnd		&	16.0	\: \: &	15.9 \: \: &	\vspace{5pt}	\\	
cdm\_randn				&	5.73	&	6.26	&	\multirow{2}{*}{2.41}		\\
Matlab R2013a randn		&	14.0	\: \: &	14.8	\: \: &	\vspace{5pt}	\\	
\botrule
\end{tabular}}
\begin{tablenotes}
\tabnote{$^{\rm a}$Median runtime of three trials of generating and aggregating $10^9$ PRNs.}
\tabnote{$^{\rm b}$ Intel\textsuperscript{\textregistered} Core$^{\rm TM}$ i7-3770K \lq{}Ivy Bridge\rq{} CPU @ 3.50GHz with 8 MB cache and 32 GB ram. Compiled using gcc 4.6.3 \& all optimization flags enabled.}
\tabnote{$^{\rm c}$Intel\textsuperscript{\textregistered} Core$^{\rm TM}$ i7-2600K \lq{}Sandy Bridge\rq{} CPU @ 3.40GHz with 8 MB cache and 16 GB ram. Compiled via gcc 4.4.3 \& all optimization flags enabled.}
\tabnote{$^{\rm d}$The PRNG provided by the non-native module \lq{}numpy\rq{}\cite{scipy} substantially outperforms the standard library module \lq{}random\rq{}, so it was compared against for benchmarking}
\end{tablenotes}
\label{Table2}
\end{table}

\section{Discussion}
Here I present a modified ziggurat algorithm that places ziggurat layers beneath a desired distribution, instead of above the desired distribution. This modification simplifies calculation of exponentially- and normally-distributed PRNs in the common case and, in-conjunction with efficient sampling of the remaining probability mass overhangs, accelerates PRN generation in all cases profiled.

The modified algorithm was implemented for two of the most common probability distributions and in common programming languages used by the scientific computing community. In principle however, the algorithm could be extended to other probability distributions and, of course, other programming languages. The modified ziggurat algorithm presented here should improve performance, relative to the traditional algorithm, for nearly all probability distributions to be generate because it simply removes computational steps in the $>98\%$ of cases when rejection sampling is unnecessary. While sampling from the overhangs could conceivably be slower in this algorithm relative to the traditional algorithm, this was not the case for the distributions sampled here and, nonetheless, rejection sampling is rare with minimal impact on overall efficiency.  

Many of the properties of ziggurat algorithms that make them the most efficient PRNGs today exploit advantages of modern architectures. Specifically, ziggurat algorithms use cached lookup tables and control flow operations that execute faster today than they would on older CPUs. Alternate algorithms may be best suited for PRNGs on computers lacking these strengths. On the other hand, this algorithm and ziggurat algorithms in general, should become more competitive as greater accuracy is desired. Implementing this algorithm to greater precision does not require modifying the code in the common case in any way; only more precise mathematical operations are needed. In contrast, inverse transform sampling algorithms, while not requiring cached lookup tables or control flow statements, generally require more terms in a polynomial expansion of the transformation function to increase accuracy \cite{Transcendental2003}. Hence, a ziggurat algorithm\rq{}s speed should be even more competitive for generating PRNs beyond 64-bit precision. Moreover, inverse transform sampling stretches inputed uniform PRNs across a wide range of values in regions where $P(x)$ is small---further reducing accuracy in regions like the tail of a probability distribution. This issue does not arise with rejection sampling, providing yet another reason to use ziggurat algorithms in high-accuracy applications. In general, the needs and computational resources of a program should be considered before choosing a PRNG. 

\section*{Acknowledgments}
I would like to thank Nezar Abdennur, Anton Goloborodko, Maxim Imakaev, and Geoffrey Fudenberg for helpful discussions and comments. This work was supported by the National Cancer Institute under grant U54CA143874. 

\vspace{12pt}
\bibliographystyle{gSCS}
\bibliography{references}

\begin{thebibliography}{10}
\providecommand{\url}[1]{\normalfont{#1}}
\providecommand{\urlprefix}{Available from: }

\bibitem{Lecuyerl1992}
L'Ecuyer~P. Testing random number generators. In: Winter Simulation Conference;
  1992. p. 305--313.

\bibitem{Marsaglia1984}
Marsaglia~G, Tsang~WW. A fast, easily implemented method for sampling from
  decreasing or symmetric unimodal density functions. SIAM Journal on
  scientific and statistical computing. 1984;\hspace{0pt}5(2):349--359.

\bibitem{Marsaglia2000}
Marsaglia~G, Tsang~WW. The ziggurat method for generating random variables.
  Journal of Statistical Software. 2000;\hspace{0pt}5(8):1--7.

\bibitem{Doornik2005}
Doornik~JA. An improved ziggurat method to generate normal random samples.
  University of Oxford. 2005;\hspace{0pt}.

\bibitem{Zhang2005}
Zhang~G, Leong~PHW, Lee~DU, Villasenor~JD, Cheung~RC, Luk~W. Ziggurat-based
  hardware gaussian random number generator. In: Field Programmable Logic and
  Applications, 2005. International Conference on. IEEE; 2005. p. 275--280.

\bibitem{Rubin2006}
Rubin~H, Johnson~BC. Efficient generation of exponential and normal deviates.
  Journal of Statistical Computation and Simulation.
  2006;\hspace{0pt}76(6):509--518.

\bibitem{Saito2008}
Saito~M, Matsumoto~M. Simd-oriented fast mersenne twister: a 128-bit
  pseudorandom number generator. In: Monte carlo and quasi-monte carlo methods
  2006. Springer; 2008. p. 607--622.

\bibitem{Thomas2007}
Thomas~DB, Luk~W, Leong~PH, Villasenor~JD. Gaussian random number generators.
  ACM Computing Surveys. 2007 Nov;\hspace{0pt}39(4):11--es;
  \urlprefix\url{http://portal.acm.org/citation.cfm?doid=1287620.1287622}.

\bibitem{DeSchryver2010}
de~Schryver~C, Schmidt~D, Wehn~N, Korn~E, Marxen~H, Korn~R. A new hardware
  efficient inversion based random number generator for non-uniform
  distributions. In: Reconfigurable Computing and FPGAs (ReConFig), 2010
  International Conference on. IEEE; 2010. p. 190--195.

\bibitem{Smith2002}
Smith~WD. How to sample from a probability distribution. 2002 Apr~[cited~2014
  Mar 24\hspace{0pt}];
  \urlprefix\url{http://scorevoting.net/WarrenSmithPages/homepage/sampling.ps}.

\bibitem{scipy}
Jones~E, Oliphant~T, Peterson~P, et~al. {SciPy}: Open source scientific tools
  for {Python}. 2001--; \urlprefix\url{http://www.scipy.org/}.

\bibitem{Transcendental2003}
Oved~I. Computing transcendental functions. 2003~[cited~2014 Mar
  24\hspace{0pt}];
  \urlprefix\url{http://math.arizona.edu/~aprl/teach/iriso/transcend.ps}.

\end{thebibliography}

\pagebreak

\begin{appendix} \appendixpage 
%\appendices
\section{Source code, Installation, \& Usage}
See \url{https://bitbucket.org/cdmcfarland/fast_prng}. The Python package fast\_prng is available for automatic installation via the Python Package Index at \url{https://pypi.python.org/pypi/fast_prng}.

\section{Demonstration of Quality}
To affirm that the above implementation is mathematically correct, a statistical test \lq\lq{}quality\_test.c\rq\rq{} was created and is provided. This script allows users to sample the raw moments of generated PRNs. The raw moments of a sample are always unbiased estimators of the raw moments of the generating distribution. Therefore, they provide a quick confirmation of the random properties of a distribution. Below is a sample output of the first five raw moments of $10^{12}$ trial PRNs:

\begin{verbatim}
Created 1000000000000 exponential distributed pseudo-random numbers...
X1: 1.000001
X2: 2.000004
X3: 6.000014
X4: 24.000048
X5: 119.999965

Created 1000000000000 standard normal distributed pseudo-random numbers...
X1: 0.000000
X2: 1.000001
X3: -0.000002
X4: 3.000009
X5: -0.000041
\end{verbatim}

Deviation of these moments from expectation should scale as $1/\sqrt{N}$, i.e. one part in $10^6$ for the above test. As this is the magnitude of deviations in the test, these results suggest that the algorithm is as precise as can be reasonably measured. 

Rounding errors were avoided by calculating values for the pre-computed lookup tables: $X$, $A$, and $f(X)$, to 128-bit precision. Afterwards, these values are rounded to 64-bit precision. Lastly, because this PRNG generates numbers deterministically from a uniform PRN generator, its sequential randomness should be as good as the underlying uniform generator, which was previously demonstrated to be excellent \cite{Saito2008}. Hence, the algorithm\rq{}s sequential randomness is excellent. 

\section{Additional modifications to the algorithm that mildly increased performance\footnote{These modifications often swap floating point operations for integer operations and exploit tendencies of compilers. Hence, they may not necessarily increase performance for all architectures/compilers.}}
\begin{enumerate}
	\item Drawing $U$ from a uniformly-distributed integer on the domain $[0, 2^{64})$, for exponential random number generation, and $[-2^{63}, 2^{63})$ for normally random number generation. This strategy of using integers rather than floating-point numbers accelerates the generation of $U$, and has been described previously \cite{Rubin2006}. Expanding the range of $U$ by $2^{64}$ requires multiplying $X$ and $f_i$ by $2^{-64}$ to retain the same output. 
	\item Sampling $i, j \in [0, 256)$ from the last 8 bits of $U$, which now resides on the domain $[0, 2^{64})$, also employed previously \cite{Marsaglia2000}. Because the last 12 bits of $U$ are squashed when multiplied by the floating-point values of $X_i$ and $f_i$ (as they have 52-bit mantissas), these bits can be used for alternate purposes without altering output in any way.  
	\item For normally-distributed PRNs, the exceptional cases (steps 4-8) were executed via a do-while loop.
	\item For exponentially-distributed PRNs, the exceptional cases were executed via a tail-recursive function. 
	\item In the small overhang boxes, values guaranteed to be outside of $P(x)$ in the upper-right half of the box: $U_x > 1 - U_y$, can be transformed to fall in the lower-left halve by swapping variables, i.e. $x \leftarrow 1 - U_y$ and $y \leftarrow U_x$. 
\end{enumerate}

\section{Modifications to the code that did not increase performance}
\begin{enumerate}
	\item Increasing $i_{\rm{max}}$ to 1024 (setting $i_{\rm{max}}$ to values that are not powers of two or are smaller than a byte would drastically slow computation).
	\item Calculating a table of $\epsilon_i$ for every overhang (Figure 2). Instead, a single, maximal possible deviation $\epsilon = \mbox{max}_{i}[ \epsilon_i ]$ was used. This also avoids caching a fourth lookup table.
	\item Using the multi-operation instruction \lq\lq{}fma\rq\rq{} present in the C standard library \lq\lq{}math.h\rq\rq{}.
	\item Generating single-precision PRNs. 
\end{enumerate}
\end{appendix}
\end{document}